Aerosol effect on the mobility of cloud droplets

# Environmental Research Letters

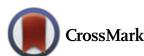

LETTER

## Aerosol effect on the mobility of cloud droplets

Ilan Koren, Orit Altaratz and Guy Dagan

Department of Earth and Planetary Sciences, Weizmann Institute, Rehovot 76100, Israel

E-mail: ilan.koren@weizmann.ac.il

## Abstract

Cloud droplet mobility is referred to here as a measure of the droplets' ability to move with ambient air. We claim that an important part of the aerosol effect on convective clouds is driven by changes in droplet mobility. We show that the mass-weighted average droplet terminal velocity, defined here as the 'effective terminal velocity' ($\eta$) and its spread ($\sigma_\eta$) serve as direct measures of this effect. Moreover, we develop analytical estimations for $\eta$ and $\sigma_\eta$ to show that changes in the relative dispersion of $\eta$ ($\varepsilon_\eta = \sigma_\eta / \eta$) can serve as a sensitive predictor of the onset of droplet-collection processes.

Droplet mobility has been previously studied in the context of movement on surfaces (Yao *et al* 2013). Here we study the mobility of cloud droplets in air. We use the term mobility to estimate how well droplets move together with the surrounding air as opposed to the deviation downward by gravity. For a given volume element that contains air and water droplets, the droplets' mobility depends heavily on the way in which water mass is distributed within the volume. The balance between drag force, buoyancy and gravity determines the droplets' terminal velocities, which are the falling velocities under zero-updraft conditions (Beard 1976, Pruppacher and Klett 1997, Reyssat *et al* 2007). Terminal velocities are inversely proportional to mobility. Both the gravity and buoyancy terms are linear and depend on the droplets' integrated (bulk) properties (total mass and density), whereas the drag force depends nonlinearly on the droplets' velocities and surface areas. This implies that a fundamental element of the droplets' dynamics depends on the shape of the size distribution.

For example, a volume element of air that contains a million 50 $\mu$m droplets has the same water mass as a volume element with one 5 mm raindrop, and so the integrated water mass of the two volumes will feel the same gravitational and buoyancy forces. Therefore in balance between all forces the two volume elements must have the same total drag force. However, the terminal velocities that are determined per-droplet will be completely different. The center of gravity of the single 5 mm raindrop falls ~30 times faster than the center of gravity of the million 50 $\mu$m droplets (relative to the surrounding air). If the air in a cloud is moving with an updraft speed of 10 m s$^{-1}$, the large raindrop will hardly move (relative to the surface) while the small droplets will have much larger mobility, as they will be pushed up at a speed of ~9.7 m s$^{-1}$. Here we show that this basic fact captures an important aerosol effect on clouds.

Aerosols—solid or liquid particles suspended in the atmosphere serve as cloud condensation nuclei or ice nuclei. Therefore, changes in aerosol concentration and properties affect the size distribution of cloud droplets (and ice crystals). Focusing on the warm part (with no ice), for a given liquid water content, a higher aerosol concentration implies more, but smaller droplets (Squires and Twomey 1966, Squires 1958). Such a change in the droplets' size distribution can affect the optical properties of the cloud (Twomey 1977, Feingold *et al* 2003) and the efficiency of collision–coalescence between droplets (Davis 1966, Warner and Twomey 1967, Albrecht 1989), as the latter depends on both the droplets' size and variance (Hsieh *et al* 2009). It therefore affects the droplets' mobility both directly by controlling the shape of the droplets' initial size distribution and indirectly by slowing down the droplets' growth driven by the collection process (redistribution to larger sizes). We show that the mass-weighted average terminal velocity, defined here as the center of gravity's 'effective terminal velocity' ($\eta$), its spread ($\sigma_\eta$) and their ratio ($\varepsilon_\eta$), can serve as direct measures of droplet mobility and therefore capture a fundamental aerosol effect on clouds. Mass-weighted average terminal velocity has been used in rain estimations (Tripoli and Cotton 1980), cirrus cloud research (Heymsfield and Iaquinta 2000, Schmitt and Heymsfield 2009) and bulk microphysics parameterizations (Straka 2009). Here we use the term 'effective terminal velocity' to highlight the fact that for a given volume element that contains droplets surrounded by air, $\eta$ predicts the vertical velocity of the integrated water mass's center of gravity (shown in equation (2) below) relative to the surrounding air. This property is inversely proportional to the water mass's ability to move with the ambient air's vertical velocity, namely its mobility.

The coupling of cloud microphysics and dynamics is at the core of convective cloud invigoration by aerosols effect (Koren *et al* 2005, Tao *et al* 2012, Altaratz *et al* 2014, Rosenfeld *et al* 2014). This effect refers to a series of feedback loops that are regulated by the properties of the droplets' size distribution. Invigoration links the increase in aerosol loading with enhanced convection in the cloud core (Khain 2009, Dagan *et al* 2015), delayed onset of precipitation (Rosenfeld 2000, Yin *et al* 2000), increased cloud water mass, larger vertical and horizontal extents (Andreae *et al* 2004, Koren *et al* 2010, Kucienska *et al* 2012, Fan *et al* 2013), longer lifetime (Teller and Levin 2006), and stronger precipitation (Heiblum *et al* 2012, Koren *et al* 2012). Evidence for cloud invigoration was first provided for deep convective clouds, suggesting that the delay in droplet freezing in polluted clouds is an important component of this effect (Khain *et al* 2005, van den Heever *et al* 2006, Rosenfeld 2008, Li *et al* 2011, Fan *et al* 2012). It has been recently shown that invigoration is a more general phenomenon that is also applicable to warm clouds and that can be viewed as an extension of the case of aerosol-limited clouds (Koren *et al* 2014, Dagan *et al* 2015).

Here we argue that the invigoration effect on warm convective clouds is driven by two main factors: (i) changes in condensation/evaporation efficiency and duration, and (ii) changes in the mobility and spread of liquid water within the cloud. The effect of the first factor can be summarized in the following way: as explained above, an increase in aerosol concentration results in more but smaller droplets. For a given liquid water content, this implies an increase in the integrated droplets' surface area, leading to higher condensation efficiency (Squires 1958, Kogan and Martin 1994, Xue and Feingold 2006, Pinsky *et al* 2013, Koren *et al* 2014, Seiki and Nakajima, 2014, Dagan *et al* 2015). Higher condensation rates drive an increase in latent heat release that further enhances updrafts (Khain *et al* 2005, Wang 2005) and the derived supersaturation. Such invigoration trends govern mostly the early stages of cloud development when the cloud is forming, the buoyancy is positive and the cloud system is more adiabatic and under supersaturation conditions. Later, the extra gained water mass will imply larger negative buoyancy due to increased water loading and larger drag force, and the larger integrated droplet surface area will enhance evaporation efficiency in the subsaturated regimes of the cloud (Xue and Feingold 2006). The second factor, specifically changes in droplet mobility, is the topic of this paper.

The terminal velocity $V_t$ increases with droplet radius $r$ and can be expressed as:

$$V_t = Qr^\kappa, \qquad (1)$$

where the terminal velocity exponent $\kappa$ gradually shifts from $\sim 2$ for small droplets (scale of 10 $\mu$m, Stokes regime, negligible Reynolds number) to $\sim 0.5$ for the turbulence raindrop range (scale of 1 mm) and $Q$ is a piecewise constant per range (Rogers and Yau 1989, Khvorostyanov and Curry 2002). We define the effective terminal velocity ($\eta$) and the effective terminal velocity spread ($\sigma_\eta$) as the mass-weighted mean terminal velocity and terminal velocity standard variation around it:

$$\eta = \frac{\int_0^\infty V_t m(r) n(r) \mathrm{d}r}{\int_0^\infty m(r) n(r) \mathrm{d}r}, \qquad (2)$$

$$\sigma_\eta = \sqrt{\frac{\int_0^\infty (V_t - \eta)^2 m(r) n(r) \mathrm{d}r}{\int_0^\infty m(r) n(r) \mathrm{d}r}} \qquad (3)$$

$\eta$, by definition, predicts the vertical displacement of the center of gravity, per unit time, under zero-updraft conditions. The non-symmetric standard deviation $\sigma_\eta$ provides an estimation of the vertical spread in altitude around the center of gravity, per unit time. Larger spread implies better mixing in the cloud, i.e., a larger likelihood of interactions between droplets from adjacent volume elements. These values can be calculated for the whole cloud, or locally per given subvolume within the cloud. The center-of-gravity view of cloud water mass has been shown to be a useful approach in reducing the large dimensionality of models or measurement outputs, replacing them with fewer variables that capture trends in the cloud's dynamics (Grabowski *et al* 2006, Koren *et al* 2009, Heiblum *et al* 2012).

The relative dispersion ($\varepsilon$) of a distribution is defined as the ratio between the standard deviation and the average. It provides information on how closely the width of the distribution evolves together with the mean. Theoretically, in a distribution with two or more free parameters, the first two moments can be completely independent and therefore $\varepsilon$ will not have any predictable value. However, many distributions in nature do show a bounded range of $\varepsilon$ values, suggesting that the distribution moments are not completely orthogonal (Tas *et al* 2015). At the limit, an invariant $\varepsilon$ value of a given distribution implies a single-parameter distribution.

A cloud's droplet size distribution can be well approximated by a gamma distribution, as shown in many *in situ* measurement and bin-microphysics modeling studies (Liu and Daum 2000, Seifert *et al* 2006, McFarquhar *et al* 2007, Xie and Liu 2009). Expressing the droplet size distribution via gamma allows us to estimate $\eta$, $\sigma_\eta$ and their ratio $\varepsilon_\eta = \sigma_\eta/\eta$ analytically as a function of the distribution's parameters for a given droplet size ($\kappa$) regime. The kernel of the distribution ($G$) in the gamma form is: $G(r) = \frac{\beta^{-\alpha}}{\Gamma(\alpha)} r^{\alpha-1} e^{-\frac{r}{\beta}}$, where $r$ is the droplet radius, $\alpha$ is a unitless shape parameter and $\beta$ is the scale parameter. For the analytical case in which $0 < G(r) < \infty$, the mean of the distribution $\bar{r}$ and the standard deviation $\sigma_r$ can be calculated directly from the distribution parameters: $\bar{r} = \alpha\beta$, $\sigma_r = \beta\sqrt{\alpha}$, and therefore the relative dispersion $\varepsilon_r = \alpha^{-0.5}$ is a function only of the shape parameter $\alpha$ (Liu and Daum 2004).

Cases in which the relative dispersion is invariant imply a constant $\alpha$ in the droplet size distribution function, which dramatically reduces the complexity in modeling the clouds using bulk schemes and representing the clouds' properties in large-scale models (Liu *et al* 2005, 2008). Several measurements and modeling studies have suggested a range of 0.3–0.4 for $\varepsilon_r$ in warm clouds (Wood 2000, Pawlowska *et al* 2006, Berg *et al* 2011, Pandithurai *et al* 2012, Tas *et al* 2012, 2015), corresponding to a range of 6–11 for $\alpha$ values.

Solving equations (2) and (3), expressing the terminal velocity as in equation (1) and using the gamma extension of the factorial expression for any real nonnegative number yields analytical expressions for the two moments and their ratio (see detailed derivations in appendix A):

$$\eta = Q\left(\frac{\bar{r}}{\alpha}\right)^\kappa \frac{(\alpha + 2 + \kappa)!}{(\alpha + 2)!}, \quad (4)$$

$$\sigma_\eta = Q\left(\frac{\bar{r}}{\alpha}\right)^\kappa \times \begin{pmatrix} ((\alpha + 2 + 2\kappa)!(\alpha + 2)! \\ -((\alpha + 2 + \kappa)!)^2)/ \\ ((\alpha + 2)!)^2) \end{pmatrix}^{1/2} \quad (5)$$

and therefore

$$\varepsilon_\eta = \begin{pmatrix} ((\alpha + 2 + 2\kappa)!(\alpha + 2)! \\ -((\alpha + 2 + \kappa)!)^2)/ \\ ((\alpha + 2 + \kappa)!)^2) \end{pmatrix}^{1/2}. \quad (6)$$

Note that the expression for $\varepsilon_\eta$ depends only on the shape parameter $\alpha$ and the terminal velocity exponent $\kappa$, and therefore for a given droplet size (specific $\kappa$ regime), a constant droplet relative dispersion ($\varepsilon_r$) implies constant $\varepsilon_\eta$. Moreover, we note that for the relevant $\kappa$ regime (0.5–2), equation (6) can be approximated linearly, well around the analytical solution of $\varepsilon_\eta$ for $\kappa = 1$ (see appendix B) as:

$$\varepsilon_\eta \cong \frac{\kappa}{\sqrt{(\alpha + 3)}}. \quad (7)$$

This shows that the effective terminal velocity variance narrows as the droplet size grows (as $\kappa$ decreases).

Such expressions allow us to estimate $\eta$, $\sigma_\eta$ and $\varepsilon_\eta$ for any $\alpha$ value and terminal velocity droplet exponent ($\kappa$) range (see appendix A). Using equations (4)–(6), the effective terminal velocity and its moments are plotted in figure 1. The upper panel shows the results for the small droplet (Stokes) regime ($\kappa = 2$) as a function of the averaged droplet radiuses, for three $\varepsilon_r$ values that cover the main regime of the measured values: 0.31 ($\alpha = 10$), 0.35 ($\alpha = 8$) and 0.41 ($\alpha = 6$). When the average radius is around 10 $\mu$m, $\eta$ and $\sigma_\eta$ are in the range of 2–3 and 1–2 cm s$^{-1}$ respectively, whereas for $\bar{r} = 60$ $\mu$m, $\eta$ and $\sigma_\eta$ are in the range of 82–112 and 46–77 cm s$^{-1}$ respectively. For a given updraft velocity, this difference in $\eta$ implies $\sim$550 m difference in the vertical location of a volume element's center of gravity after 10 min. Moreover, an over 20-fold increase in the droplets' spread around the center of gravity ($\sigma_\eta$) suggests larger mixing between droplets of adjacent vertical volumes and therefore enhanced collection. The lower panel of figure 1 shows $\eta$, $\sigma_\eta$ and $\varepsilon_\eta$ calculations for the whole droplet and raindrop range (from $\kappa = 2$ to 0.5 and for $\alpha = 8$).

In all of the above derivations, we assume analytical gamma droplet size distribution. How well does this represent a more realistic size distribution, which is the outcome of all competing nonlinear processes? And how do aerosol-driven changes in the droplets' mobility on the micro scale affect the cloud properties on the larger, macro scales? To answer these questions and to determine the significance of this effect, we employed a detailed bin-microphysics cloud model that resolves the main (deterministic and stochastic) processes that shape the droplet size distribution, as well as dynamic and entrainment schemes that affect the cloud on the macro scale (Tzivion *et al* 1987, Tzivion *et al* 1994, Reisin *et al* 1996). This allowed us to calculate $\eta$, $\sigma_\eta$ and $\varepsilon_\eta$ directly from the model outputs without using any analytical approximation (more details about the model and initial conditions can be found in appendix C).

All model runs were initialized with idealized atmospheric profiles describing a tropical moist environment (Garstang and Betts 1974). The profile included a well-mixed sub-cloud layer between the surface and $\sim$1000 m, a conditionally unstable cloud layer between 1000 and 4000 m (with 90% RH), and an overlying inversion layer (2 °C increase over 50 m). The same profile was used with three different aerosol concentrations of 25, 125, and 500 [#/cm$^3$]. The cloud forms after $\sim$30 min of simulation and lasts up to $\sim$110 min.

To describe the cloud's response to changes in aerosol loading in the clearest possible way, we present the temporal evolution of the main cloud processes and properties together (figure 2). The total liquid mass is an integral of all droplet and raindrop mass above the cloud base. The condensation/evaporation mass flux represents the rate of liquid water mass gained or lost by diffusion. The collected mass flux (CMF) is a direct measure of the redistribution within the size bins, as it measures the mass gained by the large droplet bins due to coalescence of droplets from the smaller ones. Larger aerosol loading implies more droplets and therefore larger overall surface area for more efficient condensation of the available supersaturation (figures 2(A) and (B)). On the other hand, the polluted clouds will have initially smaller droplets with smaller variance (figures 2(F) and (G)). This delays the onset of collision–coalescence processes and therefore the onset of rain (figures 2(D) and (E)). However, once rain starts, the raindrops in the polluted cloud are falling in a cloud that contains more water (figure 2(A)) in the form of smaller droplets, yielding higher collision events and therefore an efficient raindrop growth (figures 2(D) and (F)). This results in a stronger and larger amount of surface rain (figure 2(E)).

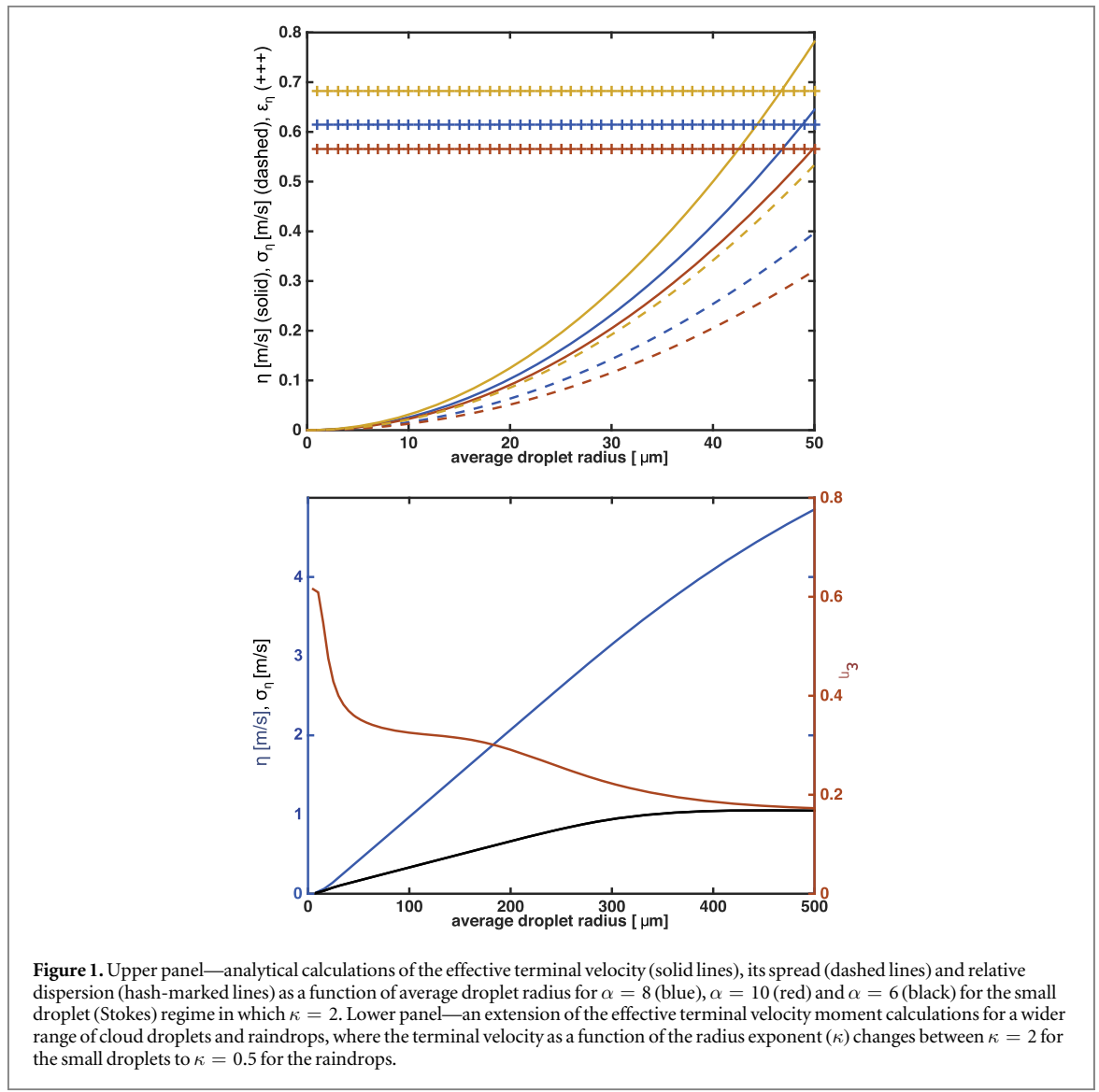

**Figure 1.** Upper panel—analytical calculations of the effective terminal velocity (solid lines), its spread (dashed lines) and relative dispersion (hash-marked lines) as a function of average droplet radius for $\alpha = 8$ (blue), $\alpha = 10$ (red) and $\alpha = 6$ (black) for the small droplet (Stokes) regime in which $\kappa = 2$. Lower panel—an extension of the effective terminal velocity moment calculations for a wider range of cloud droplets and raindrops, where the terminal velocity as a function of the radius exponent ($\kappa$) changes between $\kappa = 2$ for the small droplets to $\kappa = 0.5$ for the raindrops.

As described above, besides affecting the condensation efficiency, increased aerosol loading dramatically affects the droplet redistribution process and therefore the evolution of the shape of the droplet size distribution. This is reflected in the evolution of $\eta$, $\sigma_\eta$ and their ratio with time. Changes in the droplets' mobility imply changes in the way in which the liquid water is pushed upward by the updraft, which is a crucial influential factor in cloud development, mainly at the early stages. Changes in $\sigma_\eta$ reflect changes in the potential for interaction of droplets located in vertically adjacent volume elements. Larger $\sigma_\eta$ values mean larger spread of the droplets, by their sizes, around the location of the element's center of gravity and therefore a higher likelihood of droplet interaction between elements. In such a nonlinear system, these changes in the mobility of, and interaction between droplets (i.e. collision–coalescence) can drive dramatic changes in the cloud's evolution.

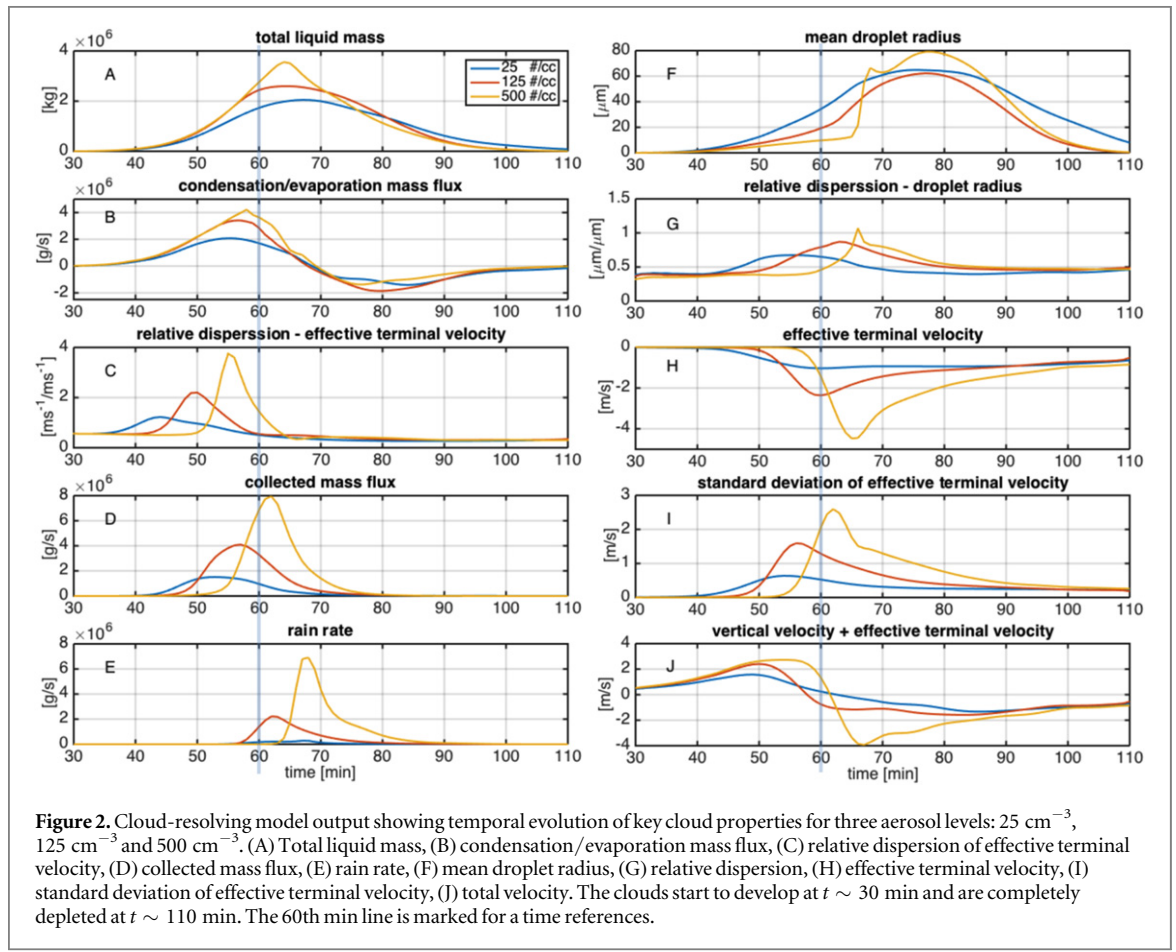

**Figure 2.** Cloud-resolving model output showing temporal evolution of key cloud properties for three aerosol levels: 25 cm$^{-3}$, 125 cm$^{-3}$ and 500 cm$^{-3}$. (A) Total liquid mass, (B) condensation/evaporation mass flux, (C) relative dispersion of effective terminal velocity, (D) collected mass flux, (E) rain rate, (F) mean droplet radius, (G) relative dispersion, (H) effective terminal velocity, (I) standard deviation of effective terminal velocity, (J) total velocity. The clouds start to develop at $t \sim 30$ min and are completely depleted at $t \sim 110$ min. The 60th min line is marked for a time references.

The early stage of cloud development, when condensation is the dominant process and the cloud can be considered closer to adiabatic, can be viewed as the cloud's build-up time. As the condensation efficiency is higher for a polluted cloud, more latent heat is released and enhances the updraft of the cloud. The cloud's updraft is coupled to changes in the liquid water mass by a few competing feedback loops. Latent heat release enhances the updraft while droplet water loading and drag force reduce it. Moreover, changes in the updraft will change the supersaturation profile of the cloud, again affecting the condensation rate. $\eta$ and $\sigma_\eta$ are determined only by the shape of the distribution (see equations (2) and (3) for the general case and equations (4) and (5) for the analytical solution of the gamma distribution case). Therefore, they can be viewed orthogonally to processes that regulate the ambient updraft. Once the updraft is known (from measurements or models), $\eta$ and $\sigma_\eta$ describe how the droplets will move with it. Figures 2(H) and (I) show the temporal evolution of the averages of $\eta$ and $\sigma_\eta$ weighted by the grid box mass. It clearly shows how the increase in aerosol concentration prolongs the duration for which $\eta$ is almost zero. This implies that the polluted cloud's updraft can push the water mass higher in the atmosphere (higher mobility) for a longer time. Later, when the collection process dominates (figure 2(D)) the polluted cloud drops fall from higher altitude through thicker cloud, with more water in the form of smaller droplets, implying more collision events at a longer path. Therefore, the droplet radius and $|\eta|$ increase more for the polluted clouds (figures 2(F) and (H)). Thus at the last stage of the clouds lifetime when the clouds are at their dissipation and rainout phase, the polluted cloud's center of gravity velocity relative to the surface (as defined by the sum of the ambient air updraft and $\eta$) becomes more negative (figure 2(J)).

The phase space spanned by the system's two characteristic velocities, i.e., the ambient vertical velocity (mean updraft weighted by the water mass, to reflect the velocity of the center of gravity as the reference) and $\eta$ (figure 3(A)), shows that they have similar scales. It also shows that the aerosol-driven delay in the onset of the collection process is reflected by an extension of the duration of negligible $\eta$ values (i.e., higher mobility of the droplets in the polluted clouds) at the early stage of the cloud's lifetime. At this stage, the total mass is still significantly smaller than the maximal one, and therefore the droplets can be pushed for a longer time, higher in the atmosphere. This accounts for a significant part of the invigoration effect. The other player in the invigoration effect, i.e., enhanced condensation in clouds that form at higher aerosol concentrations (polluted clouds), is seen as larger ambient updrafts in the first stage of the cloud's lifetime (along the upper arrow in figure 3(A)). The two warm invigoration players act together during the young stage. Later, larger collected mass by the polluted cloud's drops implies larger negative $\eta$, which infers reversed water mass movement (downward) compared to the ambient updrafts (figure 3(A), lower arrow). Note that for the polluted cloud, the main (young cloud) invigoration stage, shown as high updraft with negligible $\eta$, is well separated (almost orthogonal) from the next stage in which $\eta$ increases while the updraft decreases after consuming most of the condensation's latent heat fuel and under growing water-loading mass. The separation is less clear for the clean cloud, where the condensation efficiency is lower and collection processes start earlier. Therefore, the updraft continues to increase for a short while together with $|\eta|$. For all clouds the last (dissipation) stage is characterized by a decrease in the magnitude of $|\eta|$ and a shift in the vertical velocities to negative values driven by evaporation and rainout.

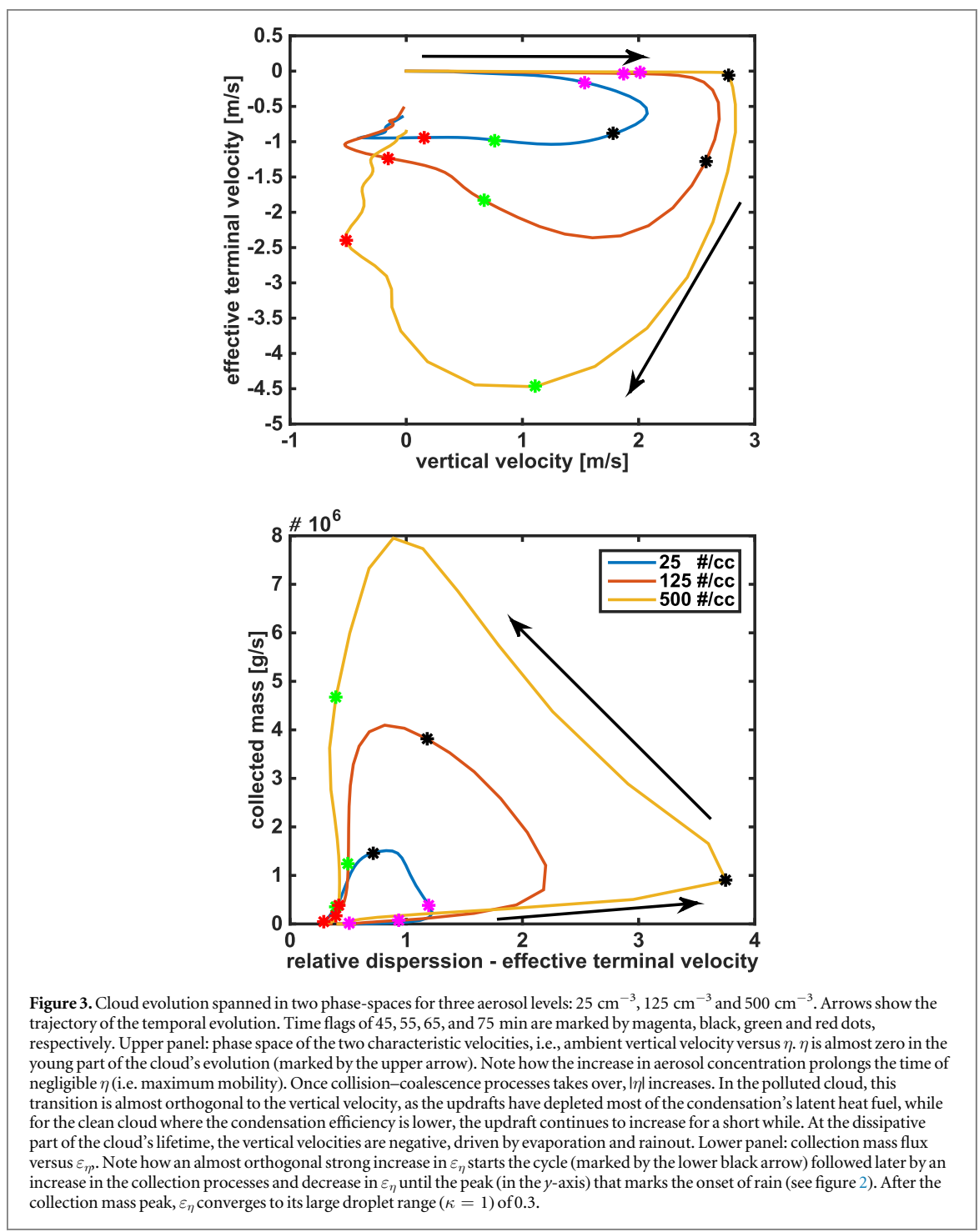

**Figure 3.** Cloud evolution spanned in two phase-spaces for three aerosol levels: 25 cm$^{-3}$, 125 cm$^{-3}$ and 500 cm$^{-3}$. Arrows show the trajectory of the temporal evolution. Time flags of 45, 55, 65, and 75 min are marked by magenta, black, green and red dots, respectively. Upper panel: phase space of the two characteristic velocities, i.e., ambient vertical velocity versus $\eta$. $\eta$ is almost zero in the young part of the cloud's evolution (marked by the upper arrow). Note how the increase in aerosol concentration prolongs the time of negligible $\eta$ (i.e. maximum mobility). Once collision–coalescence processes takes over, $|\eta|$ increases. In the polluted cloud, this transition is almost orthogonal to the vertical velocity, as the updrafts have depleted most of the condensation's latent heat fuel, while for the clean cloud where the condensation efficiency is lower, the updraft continues to increase for a short while. At the dissipative part of the cloud's lifetime, the vertical velocities are negative, driven by evaporation and rainout. Lower panel: collection mass flux versus $\varepsilon_\eta$. Note how an almost orthogonal strong increase in $\varepsilon_\eta$ starts the cycle (marked by the lower black arrow) followed later by an increase in the collection processes and decrease in $\varepsilon_\eta$ until the peak (in the *y*-axis) that marks the onset of rain (see figure 2). After the collection mass peak, $\varepsilon_\eta$ converges to its large droplet range ($\kappa = 1$) of 0.3.

Following the temporal evolution of the droplets and effective terminal velocity moments, $\varepsilon_r$ is shown to be ∼0.35 for all clouds during the young stage of the cloud's lifetime and it converges to ∼0.45 at the dissipation stage. The deviation up from the initial value occurs when the stochastic processes take over, and is reflected in the peak of the CMF (figures 2(D) and (G)). The values of the relative dispersion of the effective terminal velocity, $\varepsilon_\eta$, are remarkably concentrated around 0.56 for the young stage and around 0.3 for the mature stage of the cloud's lifetime (figure 2(C)). These values are well within the theoretical values (see figure 1, lower panel). Such behavior for all clouds suggests that indeed, similar to the droplets' relative dispersion ($\varepsilon_r$), during significant parts of the cloud's evolution, $\varepsilon_\eta$ is well bounded and close to invariant. Moreover, these values yield $\alpha \sim 6.5$ for the early stage, assuming $\kappa = 2$, and the same range for the later stage, for $\kappa = 1$.

A clear deviation in $\varepsilon_\eta$ values is shown after the early stage of the young cloud's formation when all other droplet parameters are still stable. It starts 5–10 min before any evidence of a stochastic process (figure 2(C)). Following figures 2(C), (D) and (E) reveals the timing differences between three processes: $\varepsilon_\eta$, the collection mass flux and the surface rain, all related to the timing and magnitude of the onset of the collision–coalescence processes in the cloud. This highly stochastic process enables a four orders of magnitude droplet scale growth within $\sim 10$ min, and eventually, the production of rain. It shows that $\varepsilon_\eta$ is a sensitive predictor of the onset of collection processes. The deviation from the range of theoretical values occurs much earlier than any other significant and detectable change, such as an increase in the average radius or in the CMF when the droplets are in the Stokes ($\kappa = 2$) regime. The $\eta$ integrand depends on the droplet radius to the power of $3 + \kappa$ (5 for small droplets and 3.5 for the large ones). The deviation from the theoretical value range occurs in the small droplet range of $\kappa = 2$ (figure 2(F)), which dictates the highest sensitivity of the variance ($\sigma_\eta$) to small deviations in the distribution's tail.

This sheds light on the initiation of the stochastic processes in clouds. As in many nonlinear systems, the stochastic processes are initiated by small, barely detectable perturbations around the mean that are then further amplified by positive feedback. Here, the very first collection processes shift droplets from the small bins to the very sparsely occupied large bins. This contribution to the large drop tail of the distribution increases $\sigma_\eta$ much more than it impacts $\eta$. The increase in the local spread of the droplets within a volume element around their slow-moving center of gravity implies an increase in the likelihood of interaction between droplets of vertically adjacent volume elements. Thus a positive feedback takes place, which further increases the likelihood of collisions that further enhance the spread.

Later, when the interaction between droplets of adjacent volume elements is large enough, the stochastic processes take over and a sharp increase in the CMF is seen followed by an increase in the average droplet radius and in the rain flux (figures 2(D), (E) and (F)). Figure 3(B) demonstrates the timing differences nicely when following trajectories in the $\varepsilon_\eta$ versus CMF phase space. It shows how the increase in $\varepsilon_\eta$ (marked by the lower arrow) is almost orthogonal to any change in the CMF. Later, when the CMF increases to its maximum value and then decreases, $\varepsilon_\eta$ decreases back to the theoretical value range of the larger drops, $\sim 0.3$. The structure of the trajectories in this phase space demonstrates the aerosol concentration effects. While the general trajectory structure is similar for all aerosol levels, the length and timing are different. Processes are delayed and amplified for the polluted clouds. It shows that as the aerosol level increases, $\varepsilon_\eta$ reaches its larger maximal values later in the cloud's development. Then it shows that the collection processes that started later for the polluted clouds are stronger, and that eventually $\varepsilon_\eta$ converges for all clouds to $\sim 0.3$.

In this work, we introduced the effective terminal velocity as a measure of droplet mobility. We showed that aerosol affect the shape of the droplet spectrum by shifting the initial distribution to smaller values (compared to clean clouds) with smaller variance, and delay of the collection process, while increasing its magnitude later in the cloud lifetime. This implies a strong effect on the temporal evolution of the droplets' mobility and spread as measured by $\eta$ and $\sigma_\eta$. Moreover, we showed that $|\eta|$ values are in the same range as, but often larger than the ambient vertical velocity. As the vertical movement of the cloud's center of gravity is the sum of these two velocities, there are cases in which the ambient vertical velocity is positive (updraft), but still driven by large negative $\eta$ values the cloud center of gravity will move downward (figure 3(A), during the mature stage of the two polluted cases). We argue that the effects on droplet mobility account for an important part of the aerosol effect on warm convective clouds.

We also showed that the relative dispersion of the effective terminal velocity ($\varepsilon_\eta$) can be analytically calculated under basic assumptions of droplet distribution and is close to invariant per droplet size regime. Deviation from the theoretical values provides the very first sign of the initiation of the stochastic processes by a relative increase in the $\eta$ variance that initiates a positive feedback loop by further enhancing the mixing between droplets of different sizes. This takes place earlier than the onset of any related stochastic process and therefore can serve as a sensitive predictor for it.

We note that several simplifications were done in the formation of the droplet distribution and terminal velocities as treated in the first theoretical part of the paper, to allow an analytical description of the problem. The gamma function does not necessarily represent a realistic droplet size distribution throughout the cloud evolution. In fact, $\varepsilon_\eta$ can serve as a sensitive measure for deviations from the theoretical distribution. Dividing the terminal velocity exponent ($\kappa$) to 3 size regimes is also a simplification of a challenging problem in which $\kappa$ depends on the turbulent airflow characteristics, the elevation in the atmosphere and the shape deviation from sphericity of the larger drops. Such limitations are expected whenever one tries to approximate complex processes in the cloud by smooth analytical functions. However, the numerical simulations were done using a cloud model with bin-microphysical schemes that resolves the complex interactions between droplets without assuming any distribution and accounts for turbulence. The similarity between the analytical calculations of $\eta$ and $\sigma_\eta$ (as a function of the average droplet radius) and the numerical model results suggests that the essence of the effective terminal velocity evolution in time and how it is affected by aerosols is captured by the analytical representation. Moreover, under the assumption of a constrained relative dispersion of the droplets size distribution ($\varepsilon_r$), we suggest a link between the droplets' properties and the effective terminal velocity moments (figure 1, lower panel). This can be a step forward in estimating the effective terminal velocity properties and the derived aerosol effects on the mobility of cloud droplets from measurements (*in situ* and remote sensing).

## Acknowledgments

The research leading to these results received funding from the European Research Council (ERC) under the European Union's Seventh Framework Programme (FP7/2007-2013)/ERC Grant agreement no. 306965 (CAPRI).

## Appendix A. Deriving the analytical expressions for effective terminal velocity ($\eta$), effective terminal velocity spread ($\sigma_\eta$), and their relative dispersion ($\varepsilon_\eta$)

Expressing the distribution by gamma distribution ($G(r)$) allowed us to express $\eta$, $\sigma_\eta$ and $\varepsilon_\eta$ analytically as a function of the distribution parameters for a given droplet size ($\kappa$) regime.

Writing the paper's equations (2) and (3) as:

$$\eta = \frac{\int_0^\infty V_t G(r) \rho \frac{4}{3}\pi r^3 dr}{\int_0^\infty G(r) \rho \frac{4}{3}\pi r^3 dr} \quad (A.1)$$

and

$$\sigma_\eta = \sqrt{\frac{\int_0^\infty (V_t - \eta)^2 \cdot G(r) \rho \frac{4}{3}\pi r^3 dr}{\int_0^\infty G(r) \rho \frac{4}{3}\pi r^3 dr}} \quad (A.2)$$

we note that the kernels of all integrals maintain the Euler integral of the second (gamma) form of: $r^\theta e^{-\frac{r}{\xi}}$. Using the fact that the gamma function can be viewed as an extension of the factorial function for any non-negative real number (i.e. $\Gamma(x) = (x-1)!$) yields an analytical solution for the integration range $[0\, \infty]$: $\int_0^\infty r^\theta e^{-\frac{r}{\xi}} dr = \xi^{\theta+1} \theta!$.

Using this equality and writing the terminal velocity using equation (1) ($V_t = Qr^\kappa$) allows us to solve equations (A.1) and (A.2) analytically:

$$\eta = \frac{\int_0^\infty Q r^\kappa \rho \frac{4}{3}\pi r^3 \frac{\beta^{-\alpha}}{\Gamma(\alpha)} r^{\alpha-1} e^{-\frac{r}{\beta}} dr}{\int_0^\infty \rho \frac{4}{3}\pi r^3 \frac{\beta^{-\alpha}}{\Gamma(\alpha)} r^{\alpha-1} e^{-\frac{r}{\beta}} dr}$$

$$= \frac{\int_0^\infty Q r^{\alpha+1+\kappa} e^{-\frac{r}{\beta}} dr}{\int_0^\infty r^{\alpha+2} e^{-\frac{r}{\beta}} dr}$$

$$= Q\beta^\kappa \frac{(\alpha + 2 + \kappa)!}{(\alpha + 2)!}, \quad (A.3)$$

$$\sigma_\eta = \sqrt{\frac{\left(\int_0^\infty (Qr^\kappa - \eta)^2 \rho \frac{4}{3}\pi r^3 \times \frac{\beta^{-\alpha}}{\Gamma(\alpha)} r^{\alpha-1} e^{-\frac{r}{\beta}} dr\right)/}{\left(\int_0^\infty \rho \frac{4}{3}\pi r^3 \frac{\beta^{-\alpha}}{\Gamma(\alpha)} r^{\alpha-1} e^{-\frac{r}{\beta}} dr\right)}}$$

$$= Q\beta^\kappa \left(\frac{\left((\alpha+2+2\kappa)!(\alpha+2)! -((\alpha+2+\kappa)!)^2\right)/}{\left(((\alpha+2)!)^2\right)}\right)^{1/2}. \quad (A.4)$$

Therefore $\varepsilon_\eta$, which is the ratio of the above two expressions, yields a simple expression that is a function of $\alpha$ only as shown in the paper's equation (6):

$$\varepsilon_\eta = \left(\frac{\left((\alpha+2+2\kappa)!(\alpha+2)! -((\alpha+2+\kappa)!)^2\right)/}{\left(((\alpha+2+\kappa)!)^2\right)}\right)^{1/2}.$$

In a similar manner, we can further explore the average droplet radius $\bar{r}$:

$$\bar{r} = \frac{\int_0^\infty r \frac{\beta^{-\alpha}}{\Gamma(\alpha)} r^{\alpha-1} e^{-\frac{r}{\beta}} dr}{\int_0^\infty \frac{\beta^{-\alpha}}{\Gamma(\alpha)} r^{\alpha-1} e^{-\frac{r}{\beta}} dr} = \alpha\beta. \quad (A.5)$$

This allows us to conveniently replace $\beta$ with $\bar{r}$ in equations (A.3) and (A.4), yielding the expressions shown in the paper's equations (4) and (5).

## Appendix B. A simple aproximation for $\varepsilon_\eta$

The general expression for $\varepsilon_\eta$ in the paper's equation (6) depends only on the shape parameter $\alpha$ and the terminal velocity exponent $\kappa$. The $\kappa$ values change between 2 for the small droplet (Stokes) regime ($r < 40\,\mu m$) and 0.5 for the large raindrops (Rogers and Yau 1989, Khvorostyanov and Curry 2002). Solving equation (6) for the intermediate case of droplets in the range of ∼40 μm < r < 600 μm, the terminal velocity approximated with $\kappa = 1$ yields:

$$\varepsilon_\eta(\kappa = 1) = (\alpha + 3)^{-1/2}. \quad (B.1)$$

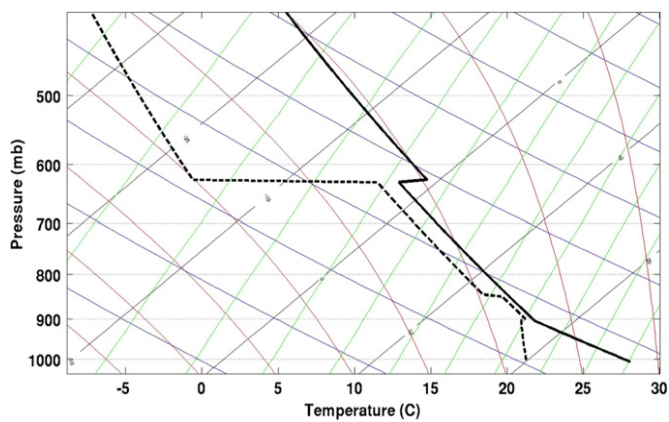

**Figure C1.** Initial atmospheric conditions used for the simulations. The solid line represents the temperature while the dashed line represents the dew point temperature.

This simple solution allows us to approximate the analytical solutions of equation (6) for the other $\kappa$ values (0.5 and 2) around it, as shown in the paper's equation (7): $\left(\varepsilon_\eta = \frac{\kappa}{\sqrt{(\alpha + 3)}}\right)$. The error of the approximation in equation (7) compared to the exact solution in equation (6) is on the order of 1% for the relevant $\alpha$ range of 6–11.

## Appendix C. Model description and initialization profile

The Tel Aviv University axisymmetric nonhydrostatic cloud model was used, with a detailed treatment of cloud microphysics (Tzivion *et al* 1994, Reisin *et al* 1996). The included warm microphysical processes were nucleation of cloud condensation nuclei, condensation and evaporation, collision–coalescence, breakup, and sedimentation. The microphysical processes were formulated and solved using a multi-moment bin method (Tzivion *et al* 1987). The model resolution was set to 50 m in both the vertical and horizontal directions, with a time step of 1 s. Convection was initiated by a warm perturbation of 3 °C at one grid point near the bottom of the domain.

The initial conditions were based on an idealized atmospheric profile describing a tropical moist environment (Garstang and Betts 1974) (figure C1). The profile includes a well-mixed subcloud layer between 0 and ∼1000 m, a conditionally unstable cloud layer between 1000 and 4000 m, and an overlying inversion layer (2 °C increase over 50 m). We assigned a dew point temperature profile that is equivalent to RH = 90% in the cloudy layer.